\documentclass[10pt,a4paper]{article}

\usepackage{a4,german,amsmath,bbm,times}

\newcommand{\cc}{\mathbbm{C}}
\newcommand{\id}{\mathbbm{1}}

\begin{document}

\bigskip\bigskip\bigskip

\noindent 
{\LARGE{\bf Multi-particle entanglement}}

\bigskip\bigskip

\noindent
{\large{J.\ Eisert and D.\ Gross}}

\bigskip

\noindent
{\it 1 Blackett Laboratory\\  
Imperial College London\\
London SW7 2BW, UK\smallskip\\
2 Institute of Physics\\
University of Potsdam\\
D-14469 Potsdam, Germany}

\bigskip\bigskip
\bigskip\bigskip

\newcommand{\scratch}[1]{{\sc Scratch:} #1}
\newcommand{\replace}[1]{{\sc Replace:} #1}

\section{Introduction}

Multi-particle entanglement is genuinely different from
entanglement in quantum systems consisting of two parts.
The prefix \emph{multi} may refer here
to quantum systems composed of a macroscopic
number of subsystems, such as the
parts of an interacting many-body system,
or it may mean merely ``three''. 
To fathom what is so different 
consider, say, a quantum system that is composed of
three qubits. Each of the qubits is thought to be held
by one of the paradigmatic distantly separated
parties. It may come as quite a surprise that
states of such composite quantum systems may contain
tri-partite entanglement, while
at the same time showing no bi-partite entanglement at all.
Such quantum states can only be generated when
all parties come together and prepare the state using
local physical devices. Whenever any
two parties group together, the state becomes separable,
and hence contains no bi-partite entanglement at all.

In this chapter, we aim at fleshing out in what ways
this multi-partite setting
is different from the situation that we
encountered earlier in this book. It is still true that
entanglement can be conceived as that property of
states that can be exploited to overcome
constraints of locality. Yet, locality refers here to
the several distinct subsystems, and we indeed
already encounter a much richer situation when 
asking questions of what states are equivalent up to
a mere local change of basis. In sharp contrast
to the bi-partite setting, there is no longer a 
natural ``unit'' of entanglement, the role that 
was taken by the maximally entangled state of 
a system of two qubits. Quite 
strikingly, 
the very concept of being maximally entangled 
becomes void. Instead, we will see that in two ways there are
``inequivalent kinds of en\-tangle\-ment''. 
We will explore some of the ramifications
of these inequivalent kinds of entanglement. 

Space limitations do not allow for a treatment of this
subject matter in full detail, yet, we aim at
``setting the coordinates'' and guiding through the extensive 
literature in this field. Our coordinate system chosen for
this chapter has the axes labeled pure and mixed states on the one hand,
entanglement in single specimens and the asymptotic setting on the other 
hand. We very briefly mention ways to detect multi-particle
entanglement, and introduce the concept of stabilizer and
graph states. Finally, we stress that multi-particle entanglement
does not only have applications in information processing as such, 
but also in metrology, for example in the context of
precision frequency standards using trapped ions. 
This chapter is emphasising the theory
of multi-particle entanglement in finite-dimensional quantum systems, 
however, we will mention 
key experimental achievements whenever possible, notably
using ion traps and purely optical systems.

\section{Pure states}\label{mpePureStates}

We will first fix one dimension in our coordinate system, and consider
multi-particle en\-tangle\-ment of \emph{pure} quantum states.  This is
the study of state vectors in a Hilbert space
\begin{equation}
        {\cal H}= {\cal H}_1\otimes {\cal H}_2\otimes ...\otimes
        {\cal H}_N
\end{equation}
of a quantum mechanical system of $N$ constituents. We will first take a
closer look at en\-tangle\-ment in single specimens of multi-partite systems, 
that is of single ``copies''. 
We will then turn to the asymptotic regime, where one asks  
questions of inconvertability when one has many identically prepared
systems at hand.

\subsection{Classifying entanglement of single specimens}

A theory of entanglement should not discriminate states that differ
only by a local operation. Here, ``local operation'' can mean merely a
change of local bases (LU operations) or, else, general local quantum
operations assisted by classical communication, that are either
required to be succesful at each instance (LOCC) or just
stochastically (SLOCC).  For each notion of locality, the questions
that have to be addressed are how many equivalence classes exist, how
are they parameterized and how can one decide whether two given states
belong to the same class?  We will briefly touch upon these problems,
limiting our attention to equivalence under LU operations and SLOCC in
turn.

For the case of LU-equivalence of bi-partite qubit states, the
\emph{Schmidt normal form}\index{Schmidt normal form} 
\begin{equation}
	\sin \theta |0,0\rangle + \cos \theta |1,1\rangle.
\end{equation}
gives a concise answer to all the above questions.
Two quantum states are LU-equivalent if and only if
their respective Schmidt normal forms coincide. All classes are parameterized
by only one real parameter: the angle $\theta$.

Some simple \emph{parameter counting} arguments show that in the case
of $N$ qubit systems the situation must be vastly more complex.
Indeed, disregarding a global phase, it takes $2^{N+1}-2$ real
parameters to fix a normalized quantum state in
$\mathcal{H}=(\cc^2)^{\otimes N}$.
The group of local unitary transformations 
$\text{SU}(2) \times \cdots \times \text{SU}(2)$
on the other hand has $3N$ real parameters. Because the set
of state vectors that are LU equivalent to a given $|\psi\rangle$ is
the same as the image of $|\psi\rangle$ under all local unitaries, the
dimension of an equivalence class cannot exceed $3N$ (it can
be less if $|\psi\rangle$ is \emph{stabilized} by a continuous
subset of the local unitaries).
Therefore, one needs at least
$2^{N+1}-3N-2$ real numbers to parameterize the sets of inequivalent
pure quantum states \cite{LindenMultiEntanglement}. Perhaps
surprisingly -- considering the rough nature of the argument -- this
lower bound turns out to be tight \cite{CarteretMultiSchmidt}. It is a
striking result that the ratio of non-local to local parameters grows
exponentially in the number of systems. In particular, the finding
rules out all hopes of a naive generalization of the Schmidt normal
form. A general pure tri-partite qubit state, say, can \emph{not} be
cast into the form
\begin{equation}
	\sin\theta |0,0,0\rangle + \cos \theta |1,1,1\rangle
\end{equation}
by the action of local unitaries \cite{PeresNoSchmidt}.

Considerable effort has been undertaken to describe the structure of
LU-equivalence classes by the use of \emph{invariants} or \emph{normal
forms} 
\cite{LindenMultiEntanglement,
CarteretMultiSchmidt,GrasslInvariants,RainsInvariants,AcinThreeSchmidt1}.
By now, even for a general multi-particle system a normal form is
known which is a generalization of the Schmidt form in the sense that
it uses a minimal number of product vectors from a factorisable
orthonormal basis to express a given state
\cite{CarteretMultiSchmidt}.
To give the reader an impression of how such generalized forms look
like, we will briefly sketch the derivation  of the simplest case,
being defined on three qubits \cite{AcinThreeSchmidt1}.
We start with a general state vector
\begin{equation}
	|\psi\rangle = \sum_{i,j,k} \alpha_{i,j,k} |i,j,k\rangle.
\end{equation}
Define two matrices $T_0$, $T_1$ by
	$(T_i)_{j,k} = \alpha_{i,j,k}$.
If we apply a unitary operator $U_1$ with matrix elements $u_{i,j}$
to the first qubit, then the matrix $T_0$ transforms according to
	$T_0' = u_{0,0} T_0 + u_{0,1} T_1$.
The algebraic constraint 
$\det(T_0')=0$ amounts to a quadratic
equation in $u_{0,1}/ u_{0,0}$  and can thus always be
fulfilled. We go on to diagonalize $T_0'$ by applying two unitaries
$U_2, U_3$ to the second and third system such that
\begin{equation}
	U_2 T_0' U_3 = 
		\left(\begin{array}{cc}
			\lambda_0& 0 \\
			0 & 0
		\end{array}\right).
\end{equation}
By absorbing phases into the definition of the basis states
$|0\rangle_1$, $|1\rangle_1$, $|1\rangle_2$, $|1\rangle_3$, we arrive
at 
\begin{equation}\label{ThreeSchmidtDecomposition}
	(U_1 \otimes U_2 \otimes U_3) 
	|\psi\rangle
	= \lambda_0 |0,0,0\rangle + \lambda_1 e^{i\phi} |1,0,0\rangle
	+ \lambda_2 |1,0,1\rangle + \lambda_3|1,1,0\rangle +
	\lambda_4|1,1,1\rangle
\end{equation}
with real coefficients $\lambda_i$. Normalization requires $\sum_i
\lambda_i^2=1$. It is shown in Ref.\
\cite{AcinThreeSchmidt1} that
$0\leq\phi\leq\pi$ can always be achieved and further, that for a
generic\footnote{In this context \emph{generic} means for all state vectors
but a set of measure zero.} state vector the form
(\ref{ThreeSchmidtDecomposition}) is unique.
In accordance with the formula we derived earlier, the normal form
depends on five independent parameters. 

How does the situation look like if one allows the local operations to
be SLOCC? An SLOCC protocol that maps a state vector $|\psi\rangle$ to
$|\phi\rangle$ with some probability of success 
consists of several rounds in 
each of which the parties perform operations on their respective
systems, possibly depending on previous measurement results. One can
think of the protocol as splitting into different \emph{branches} 
with each measurement. It should be clear that $|\psi\rangle
\rightarrow |\phi\rangle$ is possible if and only if at least
\emph{one} of these branches does the job. The effect of 
each single branch on the state vector can be described by one
\emph{Kraus operator}\index{Kraus operator} 
$A_i$ per system:
\begin{equation}\label{localFiltering}
	|\psi\rangle \mapsto (A_1\otimes\cdots\otimes A_N) |\psi\rangle.
\end{equation}
If $|\psi\rangle$ and $|\phi\rangle$ are \emph{equivalent}
under SLOCC, then the operators $A_i$ can be chosen to be invertible
\cite{DuerThreeQubits,Frank,Aki}. Note that the term
\emph{filtering operation}\index{Filtering operation} 
is used synonymously with SLOCC.

Having thus established a framework for dealing with SLOCC operations,
we can repeat the parameter counting argument from the LU case.
By simply substituting the local unitary group by
$\text{SL}(\cc^2)\times\cdots\times\text{SL}(\cc^2)$
one finds a lower bound of $2^{N+1}-6N-2$ parameters that are
necessary to label SLOCC equivalence classes of qubit systems. 
The inevitable next step would be to adopt the classification of
equivalence classes by invariants and normal forms from the LU case
to the SLOCC one. While this has indeed by done
\cite{Frank}, we are instead going
to focus on a particularly interesting special case. Indeed, for the
case of three qubits, the above estimation formula does not give a
positive lower bound for the number of parameters and therefore one
might expect that there is only a discrete set of inequivalent
classes.

Five SLOCC-inequivalent subsets of three-qubit pure
states can be identified by inspection \cite{DuerThreeQubits}.
\emph{Product vectors}\index{Product states}
\begin{equation}
	|\psi\rangle_1|\phi\rangle_2|\pi\rangle_3
\end{equation}
certainly form a class of their own because local operations can never
create entanglement between previously unentangled systems. 
For the same reason vectors of the form
\begin{equation}
	|\psi\rangle_1|\Phi\rangle_{2,3}
\end{equation}
with some non-factoring state vector $|\Phi\rangle_{2,3}$ constitute
an SLOCC-equivalence class, the class of \emph{bi-partite} entangled
states that factor with respect to the \emph{split}\index{Split}
1-23 of the set of systems.
There are three such splits (1-23, 12-3, 13-2) giving rise to three
bi-partite classes. Calling these sets \emph{equivalence classes} is
justified, because any two entangled
bi-partite pure states are equivalent under SLOCC for qubit-systems.
Finally, we are
left with the set of \emph{fully entangled}\index{fully entangled} 
vectors that admit no representation as tensor products. Do they form
a single equivalence class? It turns out that this is not the case.

To understand why this happens, we will employ an invertible SLOCC
invariant \cite{DuerThreeQubits,Schmidt}.
Any pure state can be written in the form
\begin{equation}\label{schmidtMeasure}
	|\psi\rangle = \sum_{i=1}^R	
	\alpha_i 
	|\psi_1^{i} \rangle_1\otimes ...\otimes 
	|\psi_N^{i} \rangle_N.
\end{equation}
Now let $R_{\text{min}}$ be the minimal number of product terms
needed to express $|\psi\rangle$. A moment of thought shows that this
number is constant under the action of invertible filtering operations (we
will re-visit this invariant in Section \ref{QuantifyMulti} where its
logarithm is called the \emph{Schmidt measure}\index{Schmidt measure}). 
Now consider the states vectors
\begin{eqnarray}
	|\text{GHZ}\rangle &=&
	(|0,0,0\rangle + |1,1,1\rangle)/\sqrt{2},\\
	|\text{W}\rangle &=&
	(|0,0,1\rangle + |0,1,0 \rangle + |1,0,0\rangle )/\sqrt{3}.
\end{eqnarray}
It takes a few lines \cite{DuerThreeQubits} to show that there is no
way of expressing $|\text{W}\rangle$ using only two product terms and
hence the two states cannot be converted into each other by SLOCC.
In this sense, there are two
``inequivalent forms'' of pure tri-partite entanglement of three
qubits. Notably, neither form can
be transformed into the other with any probability of success
(however, see Section \ref{mpeMixedClassification}).
Three-qubit W-states and GHZ-states 
have already been experimentally
realized, both purely
optically using postselection \cite{Wein,Zeil} and in ion
traps \cite{Blatt}. This picture is complete: any fully
entangled state is SLOCC equivalent to either $|\text{GHZ}\rangle$ or
$|\text{W}\rangle$ \cite{DuerThreeQubits}. We conclude that the three
qubits pure states are partitioned into a total of
six SLOCC equivalence classes.

We take the occassion to exemplify some concepts in multi-particle
entanglement theory by studying the properties of the GHZ and the
W state.
A simple calculation shows that after a measurement of the observable
corresponding to the Pauli matrix $X_1$ on the first qubit, both
$|\text{GHZ}\rangle$ and $|\text{W}\rangle$ collapse into a Bell state
on the systems labeled $2$ and $3$ \emph{regardless of the measurement
outcome}. Because either state is invariant under system permutations,
we can project -- with certainty -- any pair of systems into a Bell
state by performing a suitable measurement on the remaining qubit.
This property can immediately be generalized to states on more than
three systems and is known as \emph{maximal
connectedness}\index{Maximal connectedness}
\cite{BriegelPersistentEntanglement}. 
The maximum degree of entanglement of the state into which a pair can
be projected by suitable local measurements on the other parts is the
\emph{localizable entanglement}\index{Localizable entanglement}
\cite{Localizable}.

The two states behave differently, however, if a system is traced out. 
Specifically, tracing out the first qubit of the GHZ state 
will leave the remaining systems in a complete mixture. Yet, for
$|\text{W}\rangle$ we have 
\begin{eqnarray}\label{wResidualEntanglement}
	\text{tr}_1[ |\text{W}\rangle\langle\text{W}|] 
	&=& \frac13 |0,0\rangle\langle 0,0| + \frac23 |\Psi^+\rangle
	\langle\Psi^+| 
\end{eqnarray}
where $|\Psi^+\rangle=(|0,1\rangle+|1,0\rangle)/\sqrt{2}$.  
The operator in Eq.\ (\ref{wResidualEntanglement}) 
is a mixed \emph{entangled} bi-partite
state. It is in that sense, that the entanglement of
$|\text{W}\rangle$ is more \emph{robust}\index{Robustness} 
under particle loss than the
one of $|\text{GHZ}\rangle$ \cite{DuerThreeQubits}. 
Can a ``super-robust'' fully entangled three-qubit state be conceived
that leaves any pair of systems in a Bell state if the third particle
is lost? Unfortunately not, because if $|\psi\rangle$ does not factor,
$\text{tr}_1 [|\psi\rangle\langle\psi |]$ is mixed -- and in particular
not fully entangled \cite{Tangle}. This phenomenon has been
dubbed the \emph{monogamous nature}\index{Monogamy of entanglement} 
of entanglement.

\subsection{Asymptotic manipulation of multi-particle quantum states}

Needless to say, 
instead of manipulating quantum systems at the level of single 
specimens, entanglement manipulation is meaningful in the asymptotic
limit. Here, one assumes that one has many identically prepared systems
at hand, in a state $\rho^{\otimes n}$,
and aims at transforming them 
into many other identical states $\sigma^{\otimes m}$, for large $n$ and $m$,
involving collective operations. This is the asymptotic setting
that we have previously encountered in this book. 
As mentioned before, one does not require that the
target state is reached exactly, but only with an
error that is asymptotically negligible. This setting is notably different from
the one of the previous subsection, where transformations were considered for 
single specimens of multi-partite quan\-tum sys\-tems. 

It is instructive again to briefly reconsider the situation when
only two subsystems are present. In that case, the task of 
classifying different ``kinds'' of entanglement is void. Any
bi-partite entanglement of pure states is essentially equivalent to
that of an EPR-pair. One can asymptotically transform any bi-partite
state into a number of maximally entangled qubits pairs and back, 
the achievable optimal rate in this transformation being given by the 
entropy of the reduction \cite{Reversible}. 
In this sense, one can say that there is only one ingredient
to asymptotic bi-partite entanglement: this is the maximally
entangled qubit pair. Pure states can be cha\-rac\-te\-rized by the
content of this essential ingredient, giving rise to the
{\it entropy of entanglement}, which is the unique measure
of entanglement. How is this for multi-particle systems?

Again, it turns out that the situation is much more complex
than before. Before stating how the situation is like in the multi-particle
setting, let us first make the concept of {\it asymptotic
reversibility} more precise. If $\rho^{\otimes n}$ can 
be transformed under LOCC into $\sigma^{\otimes m}$ to 
arbitrary fidelity, there is no reason why $n/m$ should be
an integer. So to simplify notation, one typically
also takes non-integer yields into account. One says that 
$|\psi\rangle^{\otimes x}$ is {\it asymptotically
reducible to  $|\phi\rangle^{\otimes y}$} under
LOCC, if for all $\delta,\varepsilon>0$ there exist
natural $n,m$ such that
\begin{equation}
	\left|
	\frac{n}{m} - \frac{x}{y}
	\right|<\delta,
	\| \Psi( |\psi\rangle\langle\psi|^{\otimes n}),
	|\phi\rangle\langle\phi|^{\otimes m}
	\|_1>1-\varepsilon.
\end{equation}
Here, $\|A\|_1=\text{tr}|A|= 
\text{tr}[(A^\dagger A)^{1/2}]$ 
denotes the trace-norm of an operators $A$
as a 
distance measure, and $\Psi$ is quantum operation which is LOCC.
If both $|\psi\rangle^{\otimes x}$ can be transformed into
$|\phi\rangle^{\otimes y}$ as well as $|\phi\rangle^{\otimes y}$
into $|\psi\rangle^{\otimes x}$, the transformation is
{\it asymptotically reversible}. In the bipartite
case, it is always true that any
$|\psi\rangle$ can be transformed into
\begin{equation}
	|\psi^+\rangle^{\otimes E(|\psi\rangle\langle\psi|) },\,\,
	|\psi^+\rangle 
	= (|0,0\rangle + |1,1\rangle)/\sqrt{2}
\end{equation}
where $E( |\psi\rangle\langle\psi|) = 
S(\text{tr}_2[|\psi\rangle\langle\psi|])$
is the entropy of entanglement, and this transformation is
asymptotically reversible \cite{Reversible,BennettMulti,ReversibleVidal}. 
Such a maximally entangled qubit
pair can hence be conceived as the only essential ingredient in 
bi-partite entanglement of pure states. This holds true not only
for qubit systems, but for systems of any finite dimension,
and with small technicalities even for infinite-dimensional
systems \cite{Infinity}.

In the multi-partite setting, 
there is no longer a single essential ingredient, but
many different ones. For pure states on
${\cal H} = {\cal H}_1\otimes ... \otimes {\cal H}_N$, 
given a set of state vectors 
\begin{equation}	
	S=
	\{ |\psi_1\rangle, ..., |\psi_k\rangle\}
\end{equation} 
for some $k$, 
one may consider their {\it entanglement span} as the set
of pure states that can reversibly be generated using $S$
under asymptotic
LOCC \cite{BennettMulti}. 
In the bi-partite setting, the entanglement span is
always given by the set of all pure states (not taking
the trivial case into account where $S$ contains
only product states). 
In the multi-particle case, 
however, it is meaningful to introduce the concept of a
{\it minimal reversible entanglement generating set (MREGS)}.
An MREGS $S$ is a set of pure states such that any other state
can be generated from $S$ by means of reversible asymptotic LOCC. It
must be minimal in the sense that no set of smaller cardinality
possesses the same property
\cite{BennettMulti,LindenMREGS,VirmaniMREGS}.

After this preparation, what is now the MREGS for, say, a
tri-partite quantum system? 
The irony is that even in this relatively
simple case, no conclusive answer is known. 
Only a few states have been identified that must be contained in any
MREGS.
At first one might be tempted
to think that three different maximally entangled qubit pairs, 
shared by two systems each, already form an MREGS. This 
natural conjecture is not immediately ruled out by what we have
seen in the previous subsection: after all, we do not aim at
transforming quantum states of single specimens, but rather allow
for asymptotic state manipulation. Yet, it can be shown that 
merely to consider maximally entangled qubit pairs is not
sufficient to construct an MREGS \cite{LindenMREGS}.
What is more, even
\begin{equation}
	S= \left\{
	|\psi^+\rangle_{1,2},
	|\psi^+\rangle_{1,3},
	|\psi^+\rangle_{2,3},
	|{\text{GHZ}}\rangle
	\right\}
\end{equation}
does not suffice. 
All these pure states are {\it
inequivalent with respect to asymptotic reducibility}, 
but there are pure states that can not be
reversibly generated from these ones alone \cite{Acin}. 
So again, we see that there are 
{\it inequivalent kinds of entanglement}. 
Because we allowed for asymptotic manipulations, the present
inequivalence is even more severe than the one encountered in the last
section.

To find general means for constructing 
MREGS constitutes one of the challenging open problems 
of the field: as long as this question is generally unresolved,
the development of a ``theory of multi-particle entanglement'' in the same
way as in the bi-partite setting seems unfeasible. 
Whereas in the latter case the ``unit'' of entanglement is entirely
unambiguous -- it is the EPR-pair -- there is no substitute for it in 
sight for multi-partite systems.
This motivates after all to consider more pragmatic approaches to
grasp multi-particle entanglement.

\section{Mixed states}

\subsection{Classifying mixed state entanglement}
\label{mpeMixedClassification}

The program pursued in Section \ref{mpePureStates} -- parameterizing
all equivalence classes of states under various types of local
operations -- can in principle also be applied to mixed states
\cite{LindenNonLocalDensities, GrasslInvariants}. However, we will content
ourselves with sketching a more rough classification scheme based
\emph{separability properties} \cite{DuerMultiQubitClassification}.

At the lowest level there is the class of states that can be prepared
using LOCC alone. Its members are called 
\emph{fully separable}\index{fully separable}
and
can be written in the form
\begin{equation}
	\rho = \sum_{i} p_i
	(\rho_1^i\otimes ... \otimes \rho_N^i).
\end{equation}
Evidently, states of this kind do not contain entanglement.
Now arrange the $N$ parts of the multi-particle system in $k\leq N$
groups. We can conceive the groups as the constituents of a
$k$-partite quantum system. This \emph{coarse graining} procedure is
called forming a \emph{$k$-partite split}\index{split} of the system
and indeed, we have less
explicitly encountered that concept before in Section
\ref{mpePureStates}. 
Having set up this terminology, it is meaningful to ask with respect
to which splits a given quantum state is fully separable. Two states
belong to the same \emph{separability class}\index{separability class}
if they are separable with respect to the same splits. Clearly, being
in the same class in this sense is a necessary condition for being
equivalent under any type of local operations.  

A state is referred to as {\it
$k$-separable}, if it is fully separable considered as a state on some
$k$-partite split. By the use of this terminology, the separability
classes can be brought into a hierarchy, where $k$-separable classes
are considered to be more entangled then $l$-separable ones for $k<l$.
States that are not separable with respect to any non-trivial split
are {\it fully inseparable}. 

The number of all splits of a composite system grows exorbitantly fast
with the number $N$ of its constituents. One is naturally tempted to
reduce the complexity by identifying \emph{redundancies} in this
classification. After all, once it is established that a state is fully
separable, there is no need to consider any further splits. While such
redundancies certainly exist, pinpointing them turns out to be subtle
and indeed gives rise to one of the more peculiar results in quantum
information theory, as will be exemplified by means of our standard
example, the three-qubit system.

The five possible splits of three systems (1-2-3, 12-3, 1-23, 13-2,
123) have already been identified in Section \ref{mpePureStates}. It
is a counter-intuitive fact that there are mixed states that are
separable with respect to any bi-partite split but are not
fully separable \cite{BennettBiPartiteSeparable}. An analogous
phenomenon does not exist for pure states. The following sub-classes
of the set of
bi-separable\footnote{\emph{Bi-separable}\index{bi-separable}
means 2-separable.} states are all non-empty \cite{DuerMultiQubitClassification}.
\begin{itemize}
	\item
	\emph{1-qubit bi-separable states} with respect to the first system
	are separable for the split 1-23 but not for 12-3 or 13-2.

	\item
	\emph{2-qubit bi-separable states} with respect to the first and
	second system are separable for the split 1-23 and 2-13, but not
	for 12-3.

	\item
	\emph{3-qubit bi-separable states} are separable with respect to any
	bi-partite split but are not fully separable.
\end{itemize}
Together with the inseparable states and the fully separable ones, the
above sets constitute a complete classification of mixed three
qubit states modulo system permutations \cite{DuerMixedClassification}.

We end this subsection with a refinement of the class of inseparable states
that will play a role in the following subsection \cite{AcinThreeQubits}.  
In this paragraph, the fully separable states
are denoted by $S$, the bi-separable ones by $B$, and lastly, the set
of all mixed states including the fully inseparable ones by $F$.
Clearly, $S \subset B \subset F$ is a hierarchy of convex sets.  Now,
recall the two different classes of genuine three-qubit pure state
entanglement that were identified in Section \ref{mpePureStates}.  We
define $W$ to be the set of states that can be decomposed as a 
convex combination of
bi-separable ones and projections onto W-type vectors and finally
rename the set of all states $GHZ$.  This leaves us with a finer
partitioning $S \subset B \subset W \subset GHZ$ of the state space in
terms of convex sets. The definition suggests that the GHZ-type
vectors are in some way more entangled than the W-states -- which
until now we had no reason to suspect.\footnote{The higher Schmidt
measure of the W-state even suggest the contrary.}

In order to justify the construction, we need to employ another tool
from Section \ref{mpePureStates}: the generalized Schmidt
normal form. A pure three-particle state is SLOCC-equivalent to
the W-state if and only if its Schmidt normal form reads
\begin{eqnarray}\label{theWClass}
	\lambda_0 |0,0,0\rangle
	+ \lambda_1 |1,0,0\rangle + \lambda_2 |1,0,1\rangle
	+ \lambda_3 |1,1,0\rangle,
\end{eqnarray}
that is, if $\lambda_4=\phi=0$. 
Comparing (\ref{theWClass}) to the general form
(\ref{ThreeSchmidtDecomposition}) shows that 
$|W\rangle + \epsilon |1,1,1\rangle$ must be of GHZ-type for any 
$\varepsilon>0$.
Physically, that means that even though one cannot turn a GHZ-type
vector 
into $|W\rangle$ using SLOCC, one can \emph{approximate} it as
closely as desired. 
Therefore, we can transform states (at least approximately) from the
outer to the inner classes: $GHZ \rightarrow W \rightarrow B
\rightarrow S$ by means of non-invertible local filtering
operations. Note that invertible local operations leave the
classification of a state invariant.
As a last remark, formula (\ref{theWClass}) and the parameter
counting considerations in Section \ref{mpePureStates} show that both
the product vectors and the $W$-type vectors form a subset of measure
zero among all pure states. Notwithstanding, it can be shown 
\cite{AcinThreeQubits} that $S$ as well as $W \setminus B$ are of finite
volume in the set of mixed states.

\subsection{Methods of detection}

One way of experimentally detecting multi-particle entanglement is to 
perform a complete quantum state tomography, and to see whether the 
resulting estimated state is consistent with an entangled state.\footnote{
The question that we will only touch here is the one of 
how to {\it computationally determine whether a known
quantum state
is in one of the mentioned separability classes}. 
It turns out that already in the bi-partite case,
deciding separability is an NP-hard problem \cite{Gurvits}. 
One can nevertheless construct hierarchies of 
efficiently decidable 
sufficient criteria for a state being, say, 
fully inseparable. This is possible in a way insuring that 
every fully entangled state is necessarily detected in
some step of the hierarchy \cite{Hierarchy,Doh}. One route towards 
finding such criteria is to cast the problem 
into a polynomially constrained optimization 
problem, involving polynomials of degree three only.
This is feasible due to the fact that any 
Hermitian matrix  for which 
        $\text{tr}[M^2]=1$,
        $\text{tr}[M^3]=1$
is a matrix that satisfies
        $\text{tr}[M]=1$, 
        $ M=M^2$, 
        $M\geq 0$,
so is one that corresponds to a pure quantum state 
\cite{Hierarchy,Polynomial}. 
This can be used to parameterize the separable states from
some separability class in terms of polynomial expressions.
Relaxing the problem to hierarchies of efficiently decidable
semi-definite programs then amount to a two-way test 
of being fully inseparable \cite{Hierarchy}. 
For alternative algorithms for deciding multi-particle 
entanglement, see Refs.\ \cite{Doh,Fer,Bruss}.
}
Depending on the context, this can be a costly procedure. 
It may be 
desirable to detect entanglement without the need of aquiring full
knowledge of the quantum state.
Such an approach can be advantagous when certain types of measurements
are more accessible than others, and when one intends to detect
entanglement based on data from these restricted types of
measurements, as such insufficient to fully reconstruct the state.
This is where {\it entanglement witnesses} \index{Entanglement witness}
come into play.

An entanglement witness $A$ is an observable that is guaranteed to have a
positive ex\-pec\-ta\-tion value on the set $S$ of all separable states. So
whenever the measurement of $A$ on some quantum state $\rho$ gives a
negative result, one can be certain that $\rho$ contains some
entanglement. It is, however, important to keep in mind that witnesses
deliver only \emph{sufficient} conditions. In addition to $S$, there
might be other, non-separable states that have a positive expectation
value with respect to $A$.

We are now going to take a more systematic look at this technique and,
at the same time, generalize it from $S$ to any compact convex set $C$
in state space. To that end, we note that the set of quantum states
$\sigma$ that satisfy the equation $\text{tr}[\sigma A]=0$ for some
observable $A$ form a \emph{hyperplane} which partitions the set of
states into two half-spaces. If $C$ is contained in one of these
half-spaces, the plane is called a \emph{supporting hyperplane}
\index{supporting hyperplane} of $C$. Each half-space is
characterized by the fact that for all its respective elements 
$\sigma$ the sign of $\text{tr}[\sigma A]$ is fixed. Now, if $\rho$ is
a state contained in the half-space ``opposing'' $C$, we have, for all
$\sigma \in C$,
\begin{equation}
	\text{tr}[A\rho]<0,\,\,\,
	\text{tr}[A\sigma]\geq 0.
\end{equation}
But $\text{tr}[A\rho]$ is nothing but the expectation value of $A$
and a negative result suffices to assert that $\rho \not\in C$. 
In this way, entanglement witnesses witness entanglement.

Witnesses can be constructed for all the convex sets that appeared in
the classification of the previous subsection. For example, 
a GHZ witness is an operator that detects states that are not 
of W-type. It is not difficult to see that 
\begin{equation}
        A_{\text{GHZ}}= 
	\frac{3}{4} \id - |\text{GHZ}\rangle\langle{\text{GHZ}}|
\end{equation}
is a GHZ witness: We have that 
$\langle {\text{GHZ}}|\rho |{\text{GHZ}}\rangle\leq 3/4$
for any W-type state, and hence $\text{tr}
[A_{\text{GHZ}} \rho]\geq 0$ for any W-type state.
More generally, such witnesses can be constructed as 
$A_{\text{GHZ}} = Q-\varepsilon \id$
with an appropriate $\varepsilon>0$, where $Q\geq 0$ is a matrix that does
not have any W-type state in its kernel.
In turn, $\rho = |{\text{GHZ}}\rangle\langle{\text{GHZ}}|$ is a state
that will evidently be detected as not being of W-type. 
Similarly, a W witness is given by
\begin{equation}\label{wWitness}
	A_{\text{W}}= \frac{2}{3} 
	\id -  |{\text{W}}\rangle\langle{\text{W}}|.
\end{equation}
Needless to say, such witnesses are an especially convenient tool in the
multi-partite setting, when they are evaluated using 
only local measurements.
Just in the same way as one can choose a basis consisting of
product matrices when performing a tomographic measurement, expectation
values of witness operators can be obtained with appropriate local
measurements, using local decompositions \cite{Guehne}. 
The detection of multi-particle entanglement
using witness operators has already been experimentally
realized \cite{WeinDetect}. Indeed, one of the estimated witness
operators in the experiment was of the form given in Eq.\
(\ref{wWitness}).

\section{Quantifying multi-particle entanglement}\label{QuantifyMulti}

Entanglement measures give an answer to the question to what degree
a quantum state is entangled. Their values are typically related to
the usefulness of the state  
for quantum in\-for\-mation applications. Entanglement measures can, for
example, be related to teleportation fidelities or rates at which a
secure key can be extracted. As has been pointed out in case of the
bi-partite setting, there are two approaches to quantify
entanglement. Firstly, in the axiomatic approach, one 
specifies certain criteria that any meaningful entanglement measure must 
satisfy, and identifies functions that fulfil all these criteria.
Secondly, one may quantify a state's en\-tangle\-ment directly in terms of
rates of a certain protocol that can optimally be achieved using that
state.

In the case considered in this chapter, 
the route taken in the bi-partite setting is not accessible: 
in particular, one cannot evaluate asymptotic rates at which one can 
distill the elements of an MREGS from a given state. This would be the
analogue of
the distillable entanglement. In turn, the cost would correspond to the rates
at which one can prepare a state asymptotically starting from MREGS elements. This 
route is inaccessible; one of the reasons being that 
the MREGS are unknown. 

More pragmatically,
one can still quantify multi-particle entanglement in terms of
mean\-ing\-ful functions that are multi-particle {\it entanglement
monotones}, \index{Entanglement monotones} 
that is,  positive functions
vanishing on separable states that do not increase under LOCC,
equipped with some physical inter\-pre\-ta\-tion.
\begin{itemize}

\item The {\it Schmidt measure} \index{Schmidt measure}
$E_S$ \cite{Schmidt,Graphs}
is the logarithm of the minimal number of
products in a product decomposition
$E_S = \log_2 (R_{\text{min}})$ (see Eq.\ (\ref{schmidtMeasure})).
It provides a
classification of multi-particle entangled states and is
an entanglement monotone. In the bi-partite case, this
measure reduces to the {\it Schmidt rank},\index{Schmidt rank} 
i.e, the rank of the reduction. This measure is particularly
suitable to quantify entanglement in graph states with
many constituents.

\item Another candidate is the 
{\it global entanglement} \index{Global entanglement} 
$E_{\text{Global}}$ of Ref.\ \cite{Global}. 
This is a measure of entanglement for an 
$N$-qubit system, equipped with a Hilbert space ${\cal H} = 
({\cc}^{2})^{\otimes N}$.\footnote{This measure of entanglement is defined as
follows: Starting point is a map $f_j$
\begin{equation}\label{TheDef}
	f_j(b) |b_1,...,b_N\rangle = \delta_{b,b_j}
	|b_1,...,b_{j-1}, b_{j+1},...,b_N\rangle,
\end{equation}
where the vectors $|b_1,...,b_N\rangle$ with $b_i\in\{0,1\}$
span the Hilbert space ${\cal H}$. The right hand side of 
Eq.\ (\ref{TheDef}) is hence either zero, or the entry
$b_j$ is omitted. This map can be extended to a map ${\cc}^{2}
\otimes ({\cc}^{2})^{\otimes N}
\rightarrow 
({\cc}^{2})^{\otimes N-1}$ by linearity.
In turn, for two vectors $x,y\in 
({\cc}^{2})^{\otimes N-1}$
one may write $x= \sum_i x_i |i\rangle $ and
$y= \sum_i y_i |i\rangle $ with $0\leq i\leq 2^{N-1}$.
For a state vector $|\psi\rangle$ the quantity
\begin{equation}
	E_{\text{Global}} = \frac{4}{N}\sum_{j=1}^N
	d(f_j(0)|\psi\rangle, f_j(1)|\psi\rangle ),
\end{equation}
where $d(x,y) =\sum_{i<j} |x_i y_j - x_j y_i|^2$, is indeed
an monotone on pure states. Convex hulls of pure-state entanglement
monotones deliver then convex monotones for mixed quantum states.}

\item The {\it geometric measure of entanglement} \cite{Geometric}
makes use of a
\index{Geometric measure of entanglement}
geometric distance to the set of product states: 
\begin{equation}
	E_{\text{Geometric}}
	= \min \| |\psi\rangle\langle\psi|  -\sigma\|_2,
\end{equation}
where $\|.\|_2$ is the Hilbert-Schmidt norm, and the minimum is taken over
all product states.

\item The 
\index{tangle}
{\it tangle} \cite{Tangle}
is a measure of entanglement suitable for systems consisting
of three qubits. This measure of entanglement is based on the entanglement
of formation, or rather on the 
\index{Concurrence}
{\it concurrence}, as
\begin{equation}
	\tau(\rho) 
	= C^2(\rho_{1-23}) - C^2(\rho_{1-2})- C^2(\rho_{1-3}).
\end{equation}
Here, $C(\rho_{i-j})$ is the concurrence 
of the reduction with respect to systems labeled
$i,j$, and $C(\rho_{1-23})$ 
is the concurrence of $\rho$ in the split $1-23$.
The concurrence, in turn, is given by $C(\rho) =
\max\{0,\lambda_1^{1/2} - \lambda_2^{1/2}- \lambda_3^{1/2}-
\lambda_4^{1/2}\}$, where $\lambda_1,...,\lambda_4$ are the singular
values of $\rho\tilde\rho$, non-increasingly ordered, and $\tilde\rho=
\id\otimes \id- \rho_1\otimes \id - \id\otimes\rho_2 + \rho$. As is by
no means obvious, this quantity is invariant under permutation of the
three systems and is in fact an entanglement monotone for three-qubit
systems. 
It can be efficiently computed and applied to mixed states without the
need for taking convex hulls.

\item The {\it relative entropy of entanglement}\index{Relative entropy 
of entanglement} in the multi-partite setting 
is defined as the minimal distance
of a given state to the set of fully separable states, quantified
in terms of the quantum relative entropy \cite{RelEntMulti}.

\end{itemize}

\section{Stabilizer states and graph states}

We now turn to a specific class of multi-particle entangled states
which provides a very useful theoretical ``laboratory'':
The stabilizer formalism provides a powerful picture
for grasping a wide class of states and operations. 
Stabilizer
states are multi-qubit quantum states that play a crucial role in 
quantum information science, in
particular in the field of 
quantum error correction. It is beyond
the scope of the present chapter to give an introduction to 
the rich literature
on the stabilizer formalism. Instead, we will very briefly
introduce the very concept of a stabilizer state and 
a graph state.

Stabilizer states form a set of quantum states that allow for an
efficient description in terms of the operators they are eigenstates of.
Let us exemplify this using the familiar GHZ state on three qubits.
Recall that $X, Y$ and $Z$ denote the well-known \emph{Pauli operators}.
It is not difficult to see that the state vector 
$|{\text{GHZ}}\rangle = 
(|0,0,0\rangle+ |1,1,1\rangle)/\sqrt{2}$ is an eigenstate of 
$Z_1\otimes Z_2\otimes \id_3$, $\id_1\otimes Z_2\otimes Z_3$,
and 
$X_1\otimes X_2\otimes X_3$ to the eigenvalue $+1$.
This alone should not be too surprising. But then, the state
vector of the GHZ state is the only state vector that has this property,
up to a global phase. So instead of
explicitly writing down the GHZ state vector, we could have specified
it by saying that it is an eigenstate of $Z_1 Z_2$, $Z_2 Z_3$, 
and $X_1 X_2 X_3$.

This idea can be pushed much further -- 
and this is when the advantage of
such a formalism becomes apparent. 
The central ingredient is the \emph{Pauli group}\index{Pauli group}.
For a single system, it is given by
\begin{equation}
	G= \{ \pm \id, \pm i \id, \pm X, \pm i X, \pm Y, \pm iY, \pm Z, \pm i Z\}. 
\end{equation}
The phases
ensure that the group is actually closed under multiplication.
The Pauli group on $N$ qubits, $G_N$, in turn consists of $N$-fold tensor
products of elements of $G$. It is a basic fact from linear algebra,
that a set of $N$ operators $\{P_1,\cdots,P_N\}$ from $G_N$ allow for
a basis of common eigenvectors if they commute mutually. The key
insight lies in the observation that this basis always contains a
unique element which is a common eigenvector $|\psi\rangle$
of all $P_i$ to the eigenvalue $+1$ \cite{Gottesman}. In
other words, the operators $P_i$ \emph{stabilize} $|\psi\rangle$.
Clearly then, $|\psi\rangle$ is also stabilized by any product
of elements of $\{P_i\}_{i=1\cdots N}$. The set of all such products
forms an abelian group, the \emph{stabilizer group}\index{stabilizer
group} which is said to be \emph{generated} by the $P_i$. The vector 
$|\psi\rangle$ is the associated \emph{stabilizer state}, which is,
again, uniquely defined by the requirement
\begin{equation}
	P_i |\psi\rangle = |\psi\rangle.
\end{equation}

We have yet to make the claim precise that the stabilizer formalism
offers an efficient describtion. State vectors are usually
specified by their expansion coefficients with respect to some product
basis in Hilbert space.  By computing lower bounds of the Schmidt
measure (cf.\ Section \ref{QuantifyMulti}), e.g.,
it can be established that
there are stabilizer states that require in the order of $2^N$
non-vanishing terms when described in any product basis
\cite{Graphs,BriegelPersistentEntanglement}.  Their stabilizer group,
on the other hand, is determined by only $N$ generators.

There is an even more compact description of stabilizer states,
based on the familiar concept of a \emph{graph} $G(V,E)$ which is
specified by a set of vertices $V$ and an edge set $E$
\cite{Graphs,SchlingeWerner,Schlinge}. 
To each graph
on $N$ vertices, a stabilizer group is associated by the following
construction. We label the vertices with numbers 1 to $N$ and denote by
$N_a$ the \emph{neighbors} of the $a$-th vertex, that is, the set of
vertices directly connected to $a$. Now, to any vertex $a$, we associate an
element $K_a$ of the Pauli group  via 
\begin{equation}
	K_a = X_a \prod_{b \in N_a} Z_b.
\end{equation}
Using the fact that the relation of ``being a neighbor'' is
symmetric, one can show that the $K_a$ commute mutually and therefore
specify a unique stabilized $|G\rangle$, the \emph{graph state vector}
of $G$. As an example, consider a \emph{linear graph} on four
vertices.
It gives rise to the generators $\{X_1 Z_2, Z_1 X_2 Z_3, Z_2 X_3 Z_4,
Z_3 X_4\}$ and that the following vector is stabilized by each of them
\begin{equation}
	|{\text{Cluster}}\rangle 
	= \frac14 \bigotimes_{a=1}^4
	\big(|0\rangle Z_{a+1}+|1\rangle\big),
\end{equation}
where we have set $Z_5=\id$.
%
It is the {\it four-qubit cluster state}
\cite{BriegelPersistentEntanglement} \index{Cluster state}, an
instance
of a family of states which form the central resource for measurement
based quantum computing.  The four-qubit cluster state has recently
been prepared in an optical experiment \cite{ClusterZeilinger}.

Any stabilizer state can be brought into the form of a graph state
using only local unitaries \cite{Schlinge}.  In particular, this means
that all multi-particle entanglement properties of stabilizer states
can be described entirely in terms of properties of graphs. 
The same holds true for the effects of local Pauli measurements 
\cite{Graphs,Schlinge,Maarten2} and
\emph{Clifford} operations
\footnote{In the context of quantum information theory, a
\emph{Clifford} operation is a unitary operator that maps elements of
the Pauli group to elements of the Pauli group under conjugation.} 
\index{Clifford}
\cite{Graphs,Maarten}
on graph states. Multi-particle entanglement, for example in terms
of the Schmidt measure, can be assessed for graph states 
\cite{Graphs}. Stabilizer circuits can be simulated
computationally more cheaply when expressed in terms of graph states
\cite{Simon}
using the rules of Ref.\  \cite{Graphs}.  They also form a convenient
and physically motivated testbed to assess the question how robust
multi-particle entangled states may be under decoherence processes
\cite{DuerDecoherence,Buchleitner}. 

\section{Applications of multi-particle entangled states}

Any protocol of quantum information science 
making use of quantum systems with more than 
two constituents may be conceived as an 
application of multi-particle entanglement. 
To pinpoint the 
specifics of multi-particle entanglement
that make a certain task possible is yet less straightforward.
Multi-particle entanglement is certainly
crucial for quantum error 
correction, where the idea is to encode logical qubits into a
larger number of qubits in in a multi-particle
entangled state, as a protection against the entanglement with an 
environment beyond actual control. 
This, in John Preskill's words, to ``fight entanglement
with entanglement''. In quantum key distribution, we will encounter
several applications of multi-particle en\-tang\-le\-ment. In quantum computing,
multi-particle entanglement plays a key role. 
In measurement based computing, as we will see later, multi-particle
entangled states forms the resource. The use of multi-particle
entanglement can then even be ``monitored'' in the course of the 
computation \cite{Graphs}. 
It is, however, not yet entirely understood what exact criteria
concerning their en\-tang\-le\-ment the involved states have to fulfil 
to render an efficient classical simulation impossible.

Yet, multi-particle
entanglement does not only facilitate processing or 
transmission of information, but also allow for
applications in 
\index{Metrology}
{\it metrology} \cite{Metrology,Metrology2,WineOld,Frequency}. 
We will shortly sketch an idea to enhance the  accuracy
of the {\it estimation of
frequencies} using multi-particle
entangled states. This applies in particular
to frequency standards based on laser-cooled ions, which 
can achieve very high accuracies \cite{Frequency}. 
Starting point is to 
prepare $N$ ions that are loaded in a trap in 
some internal state with state vector $|0\rangle$.
One may then drive an atomic transition with natural frequency $\omega_0$
to a level $|1\rangle$ by applying 
an appropriate Ramsey pulse with frequency $\omega$, 
such that the ions are 
in an equal superposition of $|0\rangle$ and $|1\rangle$. 
After a free evolution for a time $t$, the probability to
find the ions in level $|1\rangle $ is given by 
\begin{equation}
	p = (1+ \cos((\omega-\omega_0)t) )/2.
\end{equation}
Given such a preparation, one finds that if one estimates the
frequency $\omega_0$ with such a scheme, the uncertainty in the
estimated value is given by
\begin{equation}
	\delta \omega_0 = (N T t)^{-1/2}.
\end{equation}
This theoretical limit, the \emph{shot-noise} limit, can in principle be
overcome when entangling the ions initially. This idea has been 
first explored in Ref.\ \cite{WineOld}, where it was suggested to prepare 
the ions in a $N$-particle GHZ state with state vector
	$|{\text{GHZ}}\rangle = 
	(|0,0,...,0\rangle + |1,1,...,1\rangle)/\sqrt{2}$.
With such a preparation, and neglecting decoherence effects, one finds an 
enhanced precision, 
\begin{equation}
	\delta \omega_0 =   (T t)^{-1/2}/N,
\end{equation}
beating the above limit by a factor of $1/\sqrt{N}$. 
Unfortunately, while the GHZ-state provides some increase in precision
in an ideal case, it is at the same time subject to decoherence
processes.
A more 
careful analysis shows that under realistic decoherence models this 
en\-hance\-ment actually disappears for the GHZ state. 
Notwithstanding these problems, the general idea of exploiting multi-particle 
entanglement to enhance frequency-measurements can be made use 
of: For example, for $N=4$ the partially entangled preparation
\begin{eqnarray}
	|\psi\rangle & = & \lambda_0 ( |0,0,0,0\rangle + |1,1,1,1\rangle)
	+ \lambda_1(|0,0,0,1\rangle + |0,0,1,0\rangle
	+|0,1,0,0\rangle \\
	&+&  |1,0,0,0\rangle + |1,1,1,0\rangle
	+ |1,1,0,1\rangle + |1,0,1,1\rangle + |0,1,1,1\rangle\nonumber\\
	&
	+ & \lambda_2 (|0,0,1,1\rangle + |0,1,0,1\rangle+ |1,0,0,1\rangle
	+ |1,1,0,0\rangle + |1,0,1,0\rangle + |0,1,1,0\rangle),\nonumber
\end{eqnarray}
can lead to an improvement of more than $6\%$, when the 
probability distribution $\lambda_0,\lambda_1,\lambda_2$ is appropriately
chosen and appropriate measurements are performed \cite{Frequency}. 
For four ions, exciting experiments have been performed in the
meantime \cite{NewWineland}, and applied for two ions to precision
spectroscopy \cite{NewWineland2}, indeed 
showing that the shot noise limit   
can be beaten with the proper use of entanglement. 

{\it Acknowledgements.} -- This work was supported by the EPSRC, the 
EU (IST-2002-38877), the DFG (Schwerpunktprogramm QIV), and the European
Research Councils (EURYI).

\renewcommand\bibname{References}

\end{document}